# Validation of artificial intelligence containing products across the regulated healthcare industries

*Author(s) and Affiliations*

Dr. David C. Higgins *
Guest Researcher, Berlin Institute of Health, Bertolt-Brecht-Platz 3, 10117 Berlin
E-mail: dave@uiginn.com

Prof. Christian Johner
Johner Institut GmbH, Reichenaustr. 39a, 78467 Konstanz
E-mail: christian.johner@johner-institut.de

* Corresponding author: Dr. David Higgins.

## Abstract

### Purpose

The introduction of artificial intelligence / machine learning (AI/ML) products to the regulated fields of pharmaceutical research and development (R&D) and drug manufacture, and medical devices (MD) and in-vitro diagnostics (IVD), poses new regulatory problems: a lack of a common terminology and understanding leads to confusion, delays and product failures. Validation as a key step in product development, common to each of these sectors including computerized systems and AI/ML development, offers an opportune point of comparison for aligning people and processes for cross-sectoral product development.

## Methods

A comparative approach, built upon workshops and a subsequent written sequence of exchanges, summarized in a look-up table suitable for mixed-teams work.

## Results

1. A bottom-up, definitions led, approach which leads to a distinction between broad vs narrow validation, and their relationship to regulatory regimes. 2. Common basis introduction to the primary methodologies for AI-containing software validation. 3. Pharmaceutical drug development and MD/IVD specific perspectives on compliant AI software development, as a basis for collaboration.

## Conclusions

Alignment of the terms and methodologies used in validation of software products containing artificial intelligence / machine learning (AI/ML) components across the regulated industries of human health is a vital first step in streamlining processes and improving workflows.

# Introduction

Regulatory affairs is a highly advanced discipline, with a decades long history, in the fields of pharmaceutical drug development and medical device (MD) and in-vitro diagnostic (IVD) technologies. Artificial intelligence (AI), in the form of machine learning (ML), is a comparatively new technology which is being trialed across multiple areas of both MD/IVD and the pharmaceutical development pipeline.

With every new technology come new potential products but also new work-practices which must be established. In the regulated fields of the human health industries this is likely to pose new and difficult problems. Further complicating matters is the fact that AI engineers typically have little to no experience of working in a regulated environment. And, medical regulators have little understanding of AI/ML technology.

There is a lack of a common understanding of the regulatory basis for products where AI is combined with medical devices and/or the development of medicinal products. This leads to a lack of compliance on the part of the AI engineers. Which in turn, leads to prolonged approval times.

In this article, we focus on the topic of validation. This is a key step, in developing a product for use, and in the underlying regulatory processes. We assigned ourselves the following question: how can validation be understood in a common manner across med-tech, pharmaceutical drug development, and AI product development?

## Materials and Methods

This comparative study is the result of a series of online workshops conducted between the authors in the Winter of 2020-21. The workshops began with written position papers on the topic of validation, authored by the respective domain experts, on the topics of ML, software engineering, MD/IVD, and pharmaceutical drug development. These positions were then discussed and debated at length before an approach to a common understanding was agreed upon. The authors then took their respective components back to their own communities of interest in order to verify their accuracy and validate their relevance to domain experts. Finally, the outputs of this process were compiled into the present article.

## Results

The language of regulatory experts sometimes appears to proliferate with formal schemas such as purpose, requirements, and specification. The difficulty we observed through our workshops is that the specifics of language differ between application domains. Through a comparative approach, we present a light-weight standardization which can be used to align the different domains thus enabling participants to develop products together, with particular focus on the emerging technology, popularly called artificial intelligence (AI), consisting largely of machine learning (ML).

In every validation process there is an *object* which is being validated, e.g. a piece of software. Since this article concerns the validation of AI contained in tools or products, and AI today is implemented primarily in software, unless otherwise specified, we will focus on software-implemented products as the object throughout. Other examples of objects which may be validated include the entire medical device and an individual machine learning model. These examples will be discussed in the relevant sections below.

Then there is a *goal*, or *target*, of the validation. This relates to construction and functionality of the object. The most common example of a validation target is: does the object achieve the product goals in terms of its intended use?

Finally, there are a set of *methods* or *practices* which are used to validate that the object conforms to the goals of its validation. Methods are goal, or target, specific and depend strongly upon the phase in the software development life-cycle during which they take place.

We break our results down into four principal sections. We begin by bringing together existing definitions of validation, based upon the various applicable regulatory standards and perspectives. From these we highlight a useful distinction, which is already partially evident in the regulatory literature, through which we resolve a key regulatory difference between USFDA and EU regulations. We then introduce an overview of the key validation methodologies applied with particular attention to methodologies used in developing software with applications in human-health specific industries. Finally, we examine validation through the perspectives of the two major legally-defined domains of application: medical/in-vitro devices (MD/IVD) and pharmaceutical industry practices. For the pharmaceutical approach, we further distinguish between drug development (R&D) and manufacturing processes.

Table 1 summarizes the main results from this section, in a translation friendly format, examining the definitions, approaches and requirements to validation across five domains of interest.

# Definition and goals of validation

We begin by looking at the common underpinning definition of validation, and the related term verification, from ISO 9000 [1]. From these definitions we develop two new terms, namely *narrow* and *broad validation*, and relate these to conflicting perspectives on validation between USFDA and EU regulations. Resolving this regulatory difference, through the introduction of these terms, is a key first result.

## Defining Validation

Many regulatory practices reference and are derived from ISO 9000. According to ISO 9000 the twin terms, verification and validation, are defined as follows:

> **Verification**: confirmation, through the provision of objective evidence, that specified requirements have been fulfilled.

> **Validation**: confirmation, through the provision of objective evidence, that the requirements for a specific intended use or application have been fulfilled.

Unfortunately, these definitions are frequently not followed. A recent IMDRF working group has highlighted this same issue [2]. Worse, the EU medical device directive (MDD)[3] even uses the term validation following two different meanings within a single sentence:

> "For devices which incorporate software or which are medical software in themselves, the software must be **validated** according to the state of the art taking into account the principles of development lifecycle, risk management, **validation** and verification."

Instead of referring to one set of definitions, the exact meaning of the term validation rather depends on the context and the point in the product life-cycle at which it is used.

## Narrow vs broad validation

As the EU medical device directive's requirement cited above makes clear, the term validation is used in two different meanings, which we describe as follows:

1. **Narrow validation**, or validation in the narrower sense:
   Here the target of the medical software validation is *limited* to evaluating whether the right product was built, i.e. the intended purpose / use is met.

   In this case, the intended purpose / use depends on and frequently also specifies the users, the context of use, and where relevant the patient population and the medical purpose. This validation typically is performed following a black box methodology and is carried out towards the end of the product development life cycle. This is validation according to the ISO 9000 definition.

2. **Broad validation**, or validation in the broader sense:
   Broad validation *includes* narrow validation but encompasses a much wider set of activities which ultimately end in narrow validation. In this case, the target of software validation is understood as a synonym for software quality assurance over the entire software development process. i.e. that the software was developed and built applying best-practices and methods.

   These validation methods also make use of the knowledge of the inner workings of the software system. Which means the software is designed and evaluated as a white box and is carried out throughout the development process. In other words: broad validation does not only analyze (test) the object (e.g. a developed software) in order to find errors or make sure that the intended use actually is met. It also contributes to the prevention of these errors right from the beginning.

The FDA's guidance document on software validation covers the entire software development process [4]. It describes activities such as requirements analysis, software architecture and software testing, but does not describe in detail how to evaluate that the intended use is met nor how to perform usability testing. In other words, whereas the EU's MDD combines both broad and narrow validation, the FDA guidance document uses the term *software validation* exclusively in the broader sense.

## Validation frequently includes verification

Importantly, software validation in the broader sense includes the software verification activities. For example, software unit tests evaluate whether a software unit performs as specified, and indeed reviews of the documentation activities. By definition such a confirmation, that specified requirements are met, is a verification activity. Also integration and software system tests, that test against given requirements, are verification activities, even though they clearly form part of software validation in the broader sense. It is important then to understand that software verification and software validation are only disjunct, as terms, when software validation is understood in the narrower sense.

Figure 1 illustrates some examples of activities undertaken as part of broad and narrow validation. Notably, both verification and narrow validation describe subsets of broad validation activities.

# Principal validation methodologies and practices

Software validation methodologies and practices describe activities which take place at various stages throughout the product development and subsequent deployment. This section provides a common methodological basis for understanding the validation of ML containing applications across multiple domains.

The product life-cycle describes different phases in the development of a product. IEC 62034 (Medical device software - Software life-cycle processes) [5] describes a framework for medical device software processes. The focus is on the stages beginning from planning and ending with a software version release and post-market activities. IEC 82304 (Health software — Part 1: General requirements for product safety) [6] complements this by further covering life-cycle phases such as installation and end-of-life disposal, however, purely for software health products. IEC 82304 incorporates validation in the narrow sense, but without giving further guidance as to methodology. The V-Model represents a software development life-cycle approach to documentation (see Figure 2) [7]. In a modern, agile development process, the entire sequence is passed through multiple times before reaching production.

A number of methodologies are of particular relevance for validation of products, destined for use in industries dedicated to human health. We highlight them in the following subsections. Their application roughly encompases a spectrum, shifting focus from broad validation towards narrow validation, i.e. from building things correctly, to having built the correct product. We begin by introducing machine learning (ML) best practices.

ML has a long history of development outside of regulated fields and, as such, has a completely different meaning for the word validation from common usage in the fields of human health. Software quality assurance (QA), which encompasses aspects of ML development, begins firmly in broad validation but also tapers to narrow validation later in the product life-cycle. Finally, user testing and clinical evaluations are necessary to observe how users will interact with a product and whether the product is both safe and effective in the real-life use context.

## Confusion in medical AI validation

One of the most important results from our workshops is to highlight that the use of the term validation in machine learning is almost completely disjunct from how the term is used in regulatory affairs. Even within the ML community there is a historic discord, since largely resolved, as to the correct use of the term validation [8]. Therefore, it is incredibly important when communicating with machine learning teams to specify which type of validation is being referenced.

Validation of machine learning (ML) models is technically understood to be an approach to algorithm tuning [9]. Validation offers an approach for choosing algorithm learning parameters, so-called hyperparameters, such that the subsequently trained model is less likely to overfit the training data. The goal of validation, for an ML practitioner, is an *unbiased* evaluation of model performance in order to demonstrate appropriate learning algorithm parameter choice [10]. ML validation, then, is categorically not a narrow validation. Rather, it is a ML-specific aspect of broad validation. From a compliance perspective, it is important to note that having carried out ML-specific validation is no guarantee of having delivered upon intended use, the correct interpretation of validation in ML is purely around having followed specific best-practices.

Figure 3 illustrates the ML view on testing vs validation. The loop in the Train and Validate cycle can be performed as many times as desired without invalidating subsequent performance evaluation on the (set-aside) Test set.

Recent developments in the medical academic literature linguistically confuse matters. In embracing ML, the medical literature has adopted the definition of the term validation as defined for statistics [11], and applied it to ML models [12, 13]. Thus, supplanting the ML-appropriate term *testing*. Whereas ML engineers validate on validation sets and test on test sets, the medical ML literature has largely undocumented approaches to learning parameter choice and refers to testing on the ML test set as validation. The use of a separately acquired test set is increasingly referred to as external validation in the medical literature. Clearly then, the medical ML literature largely eschews broad validation and uses the term validation of a ML model exclusively in the sense of a narrow validation.

## Software quality assurance and validation

Software quality assurance is the set of practices through which quality is ensured in the software industry [14]. Quality assurance (QA) consists of constructive and analytical quality assurance. *Constructive QA* has the goal to prevent errors. *Analytical QA* has the goal to detect errors in products, components, documents and processes.

Constructive QA frequently takes the form of processes, methods and procedures, and tools which prevent the introduction of faults, or defects, during the software development process.

Analytical QA is often assessed via tests, inspections and reviews, and audits. Both of these forms of QA tend to incorporate verification as a form of validation, i.e. broad validation.

Importantly, the target of the analytical QA extends to evaluating whether the constructive QA was appropriately planned, and whether the goals of the product are achieved (i.e. narrow validation).

With regards to machine learning, software QA is considered best-practice by ML engineers, but is seen as a completely separate domain from validation as applied to the model construction and potential performance. In practice this means that professional ML engineers engage in all relevant forms of broad validation by default, ie. both software QA and ML model construction best-practices, but are overly focused on the ML model construction and tend to not be involved in the achievement of, narrow validation, of entire product goals such as clinical benefit and safety.

## Usability testing and clinical evaluation

Usability testing and clinical evaluations are two approaches which are particularly relevant to products subject to human health regulations. They both form steps on the road to evaluating real-world performance of a product.

Software is tested not just in terms of code correctness but also in the context of product usability. IEC 62366-1 (Medical Devices - Part 1: Application of usability engineering to medical devices) [15] defines a set of requirements a usability engineering process must fulfill in order to achieve product usability. In particular, it focuses on potential safety issues and hazard-related use scenarios. In order to achieve a quality user interface (UI), the UI is first specified and then undergoes both formative and summative evaluations. Whereas summative evaluation is post-hoc, according to predefined goals at the end of product development, formative evaluation is carried out throughout product development and has the target to resolve usability issues before a product is released to the market. Usability testing is the method for summative evaluation and, as such, is a form of narrow validation. Other methods, such as inspections and cognitive walkthroughs are forms of broad validation.

The EU Medical Device Regulation (MDR)[16] defines clinical evaluation as a,
> "systematic and planned process to continuously generate, collect, analyse and assess the clinical data pertaining to a device in order to verify the safety and performance, including clinical benefits, of the device when used as intended by the manufacturer."

What this means is that evidence must be acquired for the purposes of evaluating the product, with a particular focus on a state of the art clinical effectiveness and safety profile. In this context, proof of performance and benefits are the proof that intended use is achieved. Thereby clinical evaluation is validation, as stated explicitly in ISO 13485 [17]. The evidence must be collected as part of an ongoing process of clinical investigations. Clinical evaluations of products do not end when a product receives certification, rather they then continue as post-market clinical follow-up, which is part of post-market surveillance. This can lead to the withdrawal of a product if the safety profile is found to have altered or is no longer state of the art. Clearly, although individual aspects of clinical investigations can contribute to broad validation, their overall focus is on narrow validation.

## The MD/IVD view on validation

Medical devices (MD) and in-vitro diagnostics (IVD) are sometimes separately regulated in law. But in practice, they follow highly overlapping development processes and are frequently evaluated in a similar manner. Particularly in the area of software as a MD/IVD (SaMD / SaIVD) the requirements related to validation are mostly identical.

## Validation of the product

The object of a MD/IVD product validation is the product as a whole i.e. the entire MD/IVD. The goal of the validation is to demonstrate clinical benefit, and achievement of intended use, alongside appropriate safety and performance. Validation, then, is built upon clinical evaluations[1] and investigations and a summative usability evaluation. The concept of usability evaluations has already been discussed in the methodologies section above.

For software as a medical device (SaMD), the clinical evaluation is part of validation in the narrower sense. ISO 13485 [17] makes clear that clinical evaluations are part of validation:

> "As part of design and development validation, the organization shall perform clinical evaluations [...]"

Clinical validation of IVDs uses slightly different language, namely: Performance evaluation and clinical evidence. It includes scientific validity, analytical performance and clinical performance. However, the overall approach remains comparable to that of a MD.

Furthermore, since IEC 62304 [5] is both a USFDA recognised consensus standard, and a EU harmonized standard it is clear that software best-practices must be followed throughout software-based MD/IVD development. It is worth noting that IEC 62304 specifically claims that it does not cover validation of the MD, even when it consists only of software. This can be interpreted as not covering narrow validation - similarly to the USFDA's software guidelines - since clearly the descriptions in IEC 62304 describe software validation in the broader sense.

---

[1] MDR requires a clinical evaluation, IVDR a performance evaluation.

## Validation activities inside the organization

In addition to the validation of the products - the MD/IVD - other elements of the manufacturers' organization must also be validated, such as:

1. Computerized systems validation.

    Software and computerized systems used in the context of the quality system must be validated, e.g. software to collect and preprocess training data, software development tools, etc. (see ISO 13485:2021 chapter 4,1.6).

2. Validation of infrastructure, measuring equipment.

    Any elements of the infrastructure, measuring equipment and tools, that might have an impact on the safety and performance of medical devices have to be validated (see ISO 13485:2021 chapters 6.3 and 7.6).

3. Process validation.

    Also processes are scope of validation such as processes to test devices, to build and distribute software, to perform post-market surveillance (see ISO 13485:2021 chapter 7.5).

Indeed, the paragraphs from ISO 13485 cited here are almost word-for-word identical to one another. Moreover, these three areas of validation are not necessarily disjunct. For example, the process of data collection, pre-processing and training has to be validated. This includes the validation of software and software tools. And some of these tools, e.g. the tool to perform static code analysis, are in turn measuring equipment.

## Device validation versus validation of machine learning models

Within the domain of MD/IVD development, the validation of a device and the validation of a model have different goals and should not be confused. Firstly, the usability of the device may still be so poor as to lead to a device which is not fit for purpose. Similarly, even a successfully validated model, i.e. a model that exactly meets the specified quality metrics such as sensitivity, may still not deliver towards the intended purpose of the device.

An illustrative example, based upon model performance issues, might be a model which should detect a disease from a blood test. If this model is part of an IVD with an intended use for mass screening, then the *specificity* is a crucial parameter, as otherwise many patients would suffer from false positive results (fear, unnecessary additional diagnostic testing or even treatments). If this test, however, is used to identify highly infectious patients, then false negative results can hardly be accepted. That is, the *sensitivity* must rather be maximized, while accepting that this will result in a proportionately higher number of false positive results.

Manufacturers must derive the model requirements from the intended purpose / use. That is, from what the device shall contribute to diagnosing, treating, preventing, predicting, monitoring or alleviating, diseases and injuries. Furthermore, the intended user profile (e.g. what the user can be expected to understand) and intended context of use (e.g. physical parameters such as brightness and noise, and social parameters such as stress and shift operation) play important roles in product validation here. These, again, come back to the issues of usability testing and clinical evaluation.

Manufacturers additionally must take into account the "state of the art" and must maximize the benefit-risk-ratio (e.g. MDR Annex I (1) [16]). This means:

1. They have to at least meet the benefit-risk-ratio of alternatives such as competitive products which **do not** use ML.
2. They have to provide evidence that the chosen model (architecture, hyperparameters) outperforms, or is at least equivalent to, other alternative ML models / architectures.

Finally, it is worth noting that ML best-practices [10] must also be applied throughout the development of a ML model for application in a MD/IVD. This includes validation in the narrower sense towards the intended use of the device using the ML model, and also validation in the broader sense as to whether engineering best-practices were applied. Naturally, this extends to include the verification definition of validation which evaluates whether the ML model meets specific requirements.

## The pharmaceutical industry view on validation

The pharmaceutical industry has two very different aspects, namely R&D and production, both of which require validation. As an industry grounded in research and development (R&D) there is a clear necessity to validate potential drug targets, investigative molecules, and ultimately the drugs themselves before bringing products to market. In addition, the manufacture of medications must follow strict quality standards which includes validation. Software has been used for decades, across these processes, with varying requirements for validation.

In the pharmaceutical industry software is primarily a tool and not the product itself. For example, in R&D it may be used to analyze data, whereas in manufacturing it may control production processes.

Historically, software applications were purchased to fill specific purposes. In this context, the goal of software validation was equivalent to validation in the narrower sense, i.e. black box testing.

As the use of software in pharma has expanded, and the impact of software on drugs increased, the pure blackbox verification (and analytical QS in general) was insufficient to provide sufficient evidence that the software does fulfill its intended use. So too has the need to customize applications, or indeed to develop new applications for internal use. In these cases, all of the activities related to software quality, which a software development company performs, have to be performed by the software developing pharmaceutical company instead. In this context, the goal of software validation in pharma today frequently covers the entire software development life-cycle and includes validation in the broader sense, which is the subject of GAMP5 - GxP [18].

## Drug development (R&D)

The drug development process as a whole is governed by local legal implementations of the International Congress on Harmonization (ICH) guidelines [19]. The core values of these guidelines are: quality (Q), safety (S) and efficacy (E). Quality refers largely, though not entirely, to the manufacturing process. Safety and efficacy are often jointly evaluated.

The legal requirements for software validation in R&D are largely enforced via an emphasis on data integrity for the drug licensing process. Good data and record management practices (GDRP) combined with a requirement that data used, to license a drug for market, follow ALCOA principles - Attributable, Legible, Contemporaneous, Original, and Accurate - oblige software used to be developed following engineering best practices [20, 21]. This data-oriented legislative approach effectively leads to a natural split in terms of software requirements.

Software used early in the research phase of R&D is frequently 'pre-regulatory' and thus has a lower burden of validation associated [22]. Such software is most often used in scientific exploration (i.e. research) and the consequent risk that its use will lead to any human harm is much lower than that of software used in the later stages of drug development and human testing.

Once the software is being used to develop actual drug candidates, however, the 'regulatory' phase begins and the legal burdens, to demonstrate the goals of ALCOA and GDRP, quickly lead to much higher requirements in terms of software validation. Software which has outputs which may be included in the eventual regulatory dossier, and indeed may steer clinical trials e.g. PK/PD dose safety estimates, should be accompanied by a full quality system (QS) approach and risk evaluation[18, 23, 24].

From the perspective of scientists, particularly statisticians and data scientists, involved in the R&D process software validation requirements are somewhat opaque. This despite the fact that many of them are required to develop their own software. A good rule, in this case, is that software engineering best-practice methodologies should always be followed, i.e. broad validation. The complete requirements in terms of validation, also in the narrow sense, depend on an evaluation of the potential for human harm, and for large cost overruns due to failed development paths. Artificial intelligence, as a form of software, follows the same processes and requirements as any other software used in R&D.

## Manufacturing automation

Pharmaceutical manufacturing automation largely follows GAMP5 - GxP [18] standards. Here the entire software system in the intended hard- and software environments is the primary goal for validation. ISO/IEC 25010 [25] proposes eight quality attributes which must be validated in order to ensure software product quality: functionality (e.g. suitability, completeness), performance, compatibility, usability, reliability, security, maintainability and portability.

Particularly important protocols to be followed in manufacturing automation are installation, operational and performance qualification (IQ, OQ and PQ respectively). The validation methods include: installation tests, load tests, compatibility with neighboring systems, and blackbox evaluation methods.

The GAMP5 - GxP model delineates how quality assurance responsibilities have to be split between the organization developing the software and the organization that installs, configures and uses the software. Typically, the software development company will be responsible for verification methods whereas the pharmaceutical manufacturer will be responsible for validation that the product functions as intended in the context of the assembly line (i.e. narrow validation). Additionally, the pharma company typically installs, configures, parametrizes the software, sometimes even adds scripts or calculations. All this has to follow protocols, i.e. constructive QS, and is broad validation.

## Discussion

With every new technology one of the key early milestones en route to achieving widespread usage is the development of the map of how the new technology is applied in already existing domains. Due to the timelines and engineering overheads involved, in domains in which regulatory compliance is an issue, a comprehensive roadmap is even more important in industries focused on human health. This is why we have presented a comparative study which no longer treats each domain in isolation, but also attempts to establish a common language and draws direct relations between the procedures used in each.

Broad vs narrow validation was a complicated topic even in our workshop setting. Why can they not be disjunct? This in particular is one of the sources of confusion when talking about validation. ISO 9001 talks about validation specifically in the narrower sense. Whereas many of the regulators are concerned with broad validation, and then it is unclear as to whether they also include narrow validation in their definitions or not. Therefore, we make narrow validation a subset of broad validation and then use this terminology to highlight the cases in which narrow is omitted or excluded.

In parallel to the debates in the regulatory sciences about the precise meaning of validation are debates in the academic world about terminology. As a result of this, ML engineers, who are new to regulated technology development, are happy to ignore the conversation, and subsequently end up building products which do not deliver upon the narrow validation target of performance upon intended use. One of the main goals of this article is to address this gap.

In this article we have deliberately provided less depth of detail on GAMP5 - GxP. Our reasoning there is two-fold. Firstly, AI is primarily used to make either decisions or predictions. Therefore, a lot of the more interesting applications of AI are going to be involved in R&D and in clinical trial selections. These will likely be *regulated* as MD/IVDs, although we accept that they may additionally require a GxP approach in order to gain industry support. Secondly, and more importantly, we have observed the direction of travel of ML engineers. They begin by needing to understand that they are in a regulated domain and that their work is based on the input (e.g. intended purpose specification) and that their "validated output" is necessary but not sufficient to prove that the input requirements are met. They do this by following industry-specific best-practices; subsequently, they become involved in process management, e.g. QS; and, only much later do any of them understand the specifics of MD/IVD vs GxP. We see the current article as a starting point, on this path, not the end of all discussions.

Our goal in this article was to address the question, what is validation? And, how can validation be understood in a common manner across med-tech, pharma and AI? A natural next step, from this article, is to present its contents to new cross-functional teams working together to develop AI solutions which occupy the emerging cross-over space between pharma and medtech. A previous article in a similar vein [26], by one of the authors, has seen considerable success within one of the top-3 pharma companies (personal communications). By bringing cross-functional teams together, around a topic such as this, we improve their ability to successfully develop products together, and hopefully also to imagine the opportunities for entirely new categories of solutions for the future.

# Conclusion

Validation is a hugely important process across the regulated healthcare industries. With the advent of AI/ML there are clear gaps in knowledge along with inconsistencies, across the system, as to how to validate products containing these new technologies. We adopt the terms, broad and narrow validation, and use them to align the validation processes across the industries of pharmaceutical R&D, manufacturing and medical device and in-vitro diagnostics.

# Funding Statement

Neither author has received any funding towards the completion of this study.

# Conflict of Interest Statement

David Higgins derives some of his income from advising startups on their technology strategies. The current article is not expected to lead to any changes in his income.

Christian Johner owns the Johner Institute that consults medical device and IVD manufacturers. It educates these manufacturers as well as notified bodies and has released an AI guideline that is used by manufacturers and notified bodies.

# Author Contributions

DH and CJ contributed equally to this work. The original workshops were conducted jointly.

The conception of the article was a joint effort, as was writing, editing and final submission.

# References

1. ISO 9000:2015 Quality management systems — Fundamentals and vocabulary.
2. IMDRF AIMD Working Group (2021) Machine Learning-enabled Medical Devices—A subset of Artificial Intelligence-enabled Medical Devices: Key Terms and Definitions. 16
3. (2007) Council Directive 93/42/EEC of 14 June 1993 concerning medical devices.
4. (2002) General Principles of Software Validation; Final Guidance for Industry and FDA Staff. 47
5. (2006) IEC 62304:2006 Medical device software — Software life cycle processes.
6. (2016) IEC 82304-1:2016 Health software — Part 1: General requirements for product safety.
7. McHugh M, McCaffery F, Casey V (2012) Barriers to Adopting Agile Practices When Developing Medical Device Software. In: Mas A, Mesquida A, Rout T, O'Connor RV, Dorling A (eds) Softw. Process Improv. Capab. Determ. Springer, Berlin, Heidelberg, pp 141–147
8. Brownlee J (2017) What is the Difference Between Test and Validation Datasets? Mach. Learn. Mastery
9. Russell S, Norvig P (2009) Artificial Intelligence: A Modern Approach (3rd edition). Addison Wesley, Boston Columbus Indianapolis New York San Francisco Upper Saddle River Amsterdam, Cape Town Dubai London Madrid Milan Munich Paris Montreal Toronto Delhi Mexico City Sao Paulo Sydney Hong Kong Seoul Singapore Taipei Tokyo
10. Higgins DC (2021) OnRAMP for Regulating Artificial Intelligence in Medical Products. Adv Intell Syst 3:2100042
11. Altman DG, Royston P (2000) What do we mean by validating a prognostic model? Stat Med 19:453–473
12. Kim DW, Jang HY, Kim KW, Shin Y, Park SH (2019) Design Characteristics of Studies Reporting the Performance of Artificial Intelligence Algorithms for Diagnostic Analysis of Medical Images: Results from Recently Published Papers. Korean J Radiol 20:405–410
13. Park SH, Kressel HY (2018) Connecting Technological Innovation in Artificial Intelligence to Real-world Medical Practice through Rigorous Clinical Validation: What Peer-reviewed Medical Journals Could Do. J Korean Med Sci 33:e152
14. Spillner A, Linz T, Schaefer H (2014) Software Testing Foundations: A Study Guide for the Certified Tester Exam, 4th edition. Rocky Nook, Santa Barbara, CA
15. (2015) IEC 62366-1:2015 Medical devices — Part 1: Application of usability engineering to medical devices.
16. (2017) Regulation (EU) 2017/745 of the European Parliament and of the Council of 5

April 2017 on medical devices, amending Directive 2001/83/EC, Regulation (EC) No 178/2002 and Regulation (EC) No 1223/2009 and repealing Council Directives 90/385/EEC and 93/42/EEC (Text with EEA relevance. ).
17. (2016) ISO 13485:2016 Medical Devices - Quality management systems - Requirements for regulatory purposes.
18. (2008) GAMP 5 Guide: Compliant GxP Computerized Systems. ISPE
19. ICH Official web site : ICH. https://www.ich.org/. Accessed 22 Oct 2020
20. (2016) WHO Guidance ON GOOD DATA AND RECORD MANAGEMENT PRACTICES.
21. (2020) Data Integrity and Compliance With Drug CGMP Questions and Answers Guidance for Industry. FDA
22. Pitman A (2019) Mathematical and Statistical Skills in the Biopharmaceutical Industry: A Pragmatic Approach. Chapman and Hall/CRC
23. (2018) Guidance for Industry - COMPUTERIZED SYSTEMS USED IN CLINICAL TRIALS. FDA
24. (2020) 21 CFR Part 11, Electronic Records; Electronic Signatures - Scope and Application. FDA
25. (2011) ISO/IEC 25010:2011 Systems and software engineering — Systems and software Quality Requirements and Evaluation (SQuaRE) — System and software quality models.
26. Higgins D, Madai VI (2020) From Bit to Bedside: A Practical Framework for Artificial Intelligence Product Development in Healthcare. Adv Intell Syst 2000052
27. (2021) Proposal for a REGULATION OF THE EUROPEAN PARLIAMENT AND OF THE COUNCIL LAYING DOWN HARMONISED RULES ON ARTIFICIAL INTELLIGENCE (ARTIFICIAL INTELLIGENCE ACT) AND AMENDING CERTAIN UNION LEGISLATIVE ACTS.
28. (2020) ISO/IEC TR 24028:2020 Information technology — Artificial intelligence — Overview of trustworthiness in artificial intelligence.
29. (2020) ISO/IEC TR 29119-11:2020 Software and systems engineering — Software testing — Part 11: Guidelines on the testing of AI-based systems.
30. Health C for D and R (2021) Good Machine Learning Practice for Medical Device Development: Guiding Principles. FDA
31. Johner PDC (2022) AI Guideline.
32. 21 CFR Part 820 Subpart G -- Production and Process Controls.
33. AAMI TIR36:2007 (AAMI TIR 36:2007) - Validation of software for regulated processes.
34. (2017) ISO/TR 80002-2:2017 Medical device software — Part 2: Validation of software for medical device quality systems.
35. (2017) Regulation (EU) 2017/746 of the European Parliament and of the Council of 5 April 2017 on in vitro diagnostic medical devices and repealing Directive 98/79/EC and Commission Decision 2010/227/EU (Text with EEA relevance. ).
36. Guidance document Medical Devices - Clinical investigation, clinical evaluation - Guidelines on Clinical investigations: a guide for manufacturers and notified bodies - MEDDEV 2.7/4.
37. 21 CFR Part 211 -- Current Good Manufacturing Practice for Finished Pharmaceuticals.
38. 21 CFR Part 210 -- Current Good Manufacturing Practice in Manufacturing, Processing, Packing, or Holding of Drugs; General.
39. (2001) Directive 2001/83/EC of the European Parliament and of the Council of 6 November 2001 on the Community code relating to medicinal products for human use.

# Figures

**Fig. 1.** The definition of validation varies according to the application. To bring clarity, we describe broad validation as the all-encompassing validation best-practices stemming from quality management activities. Whereas, narrow validation involves only a subset of these activities.

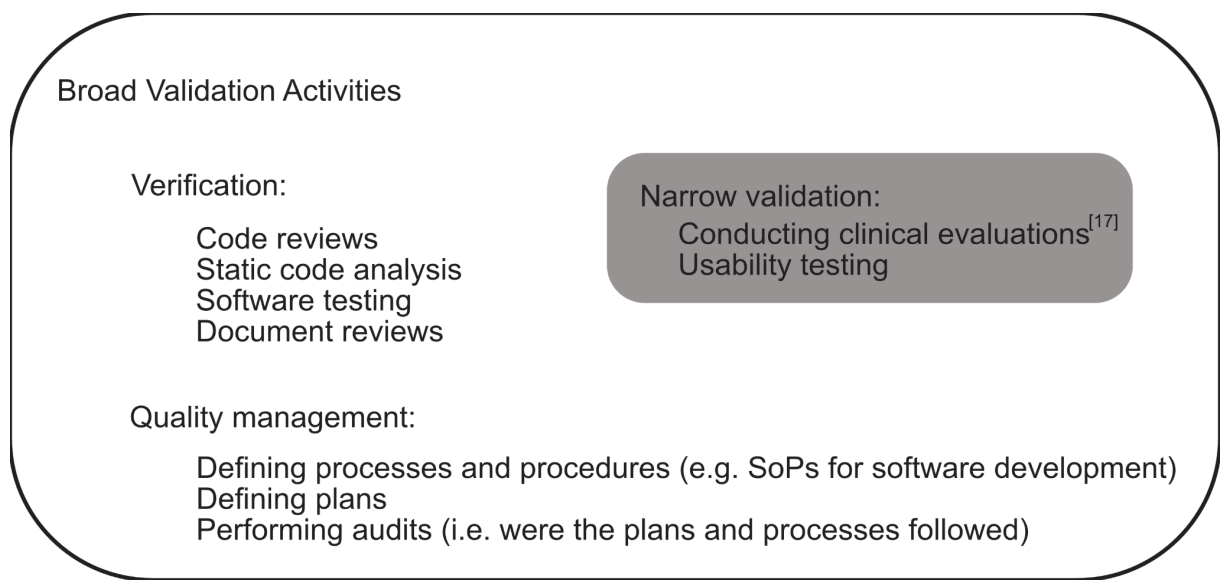

**Fig. 2.** The phases of the software development life-cycle in a V-model. For every development phase, there is a corresponding testing phase. In agile development these phases are passed through several times.

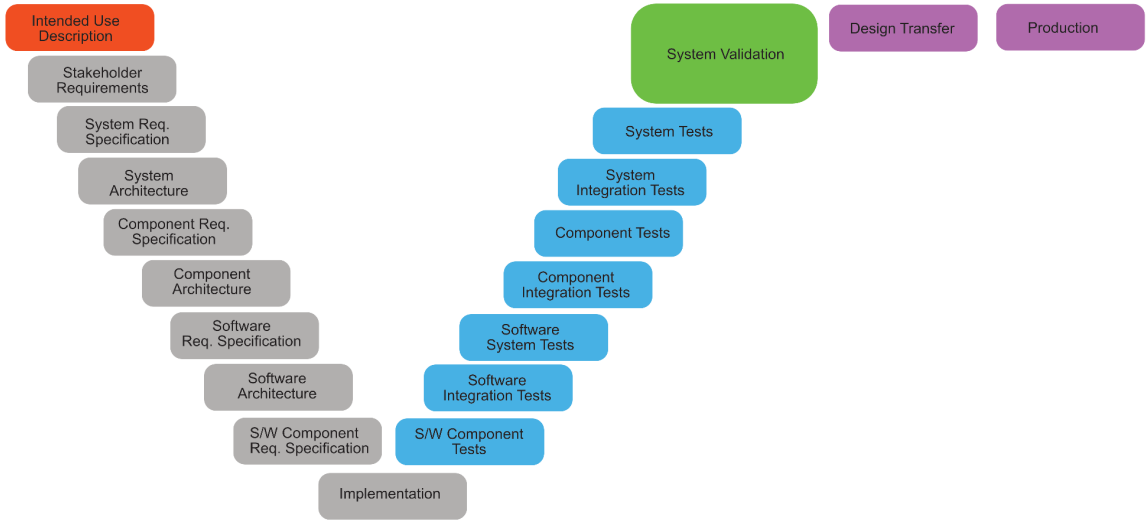

**Fig. 3.** The machine learning (ML) view on Testing vs Validation. Data is split, and locked, into separate datasets before any models are trained. Importantly, the Train-and-Validate loop can be performed as many times as desired, in order to select model hyperparameters, without biasing the subsequent performance evaluation on the Test set.

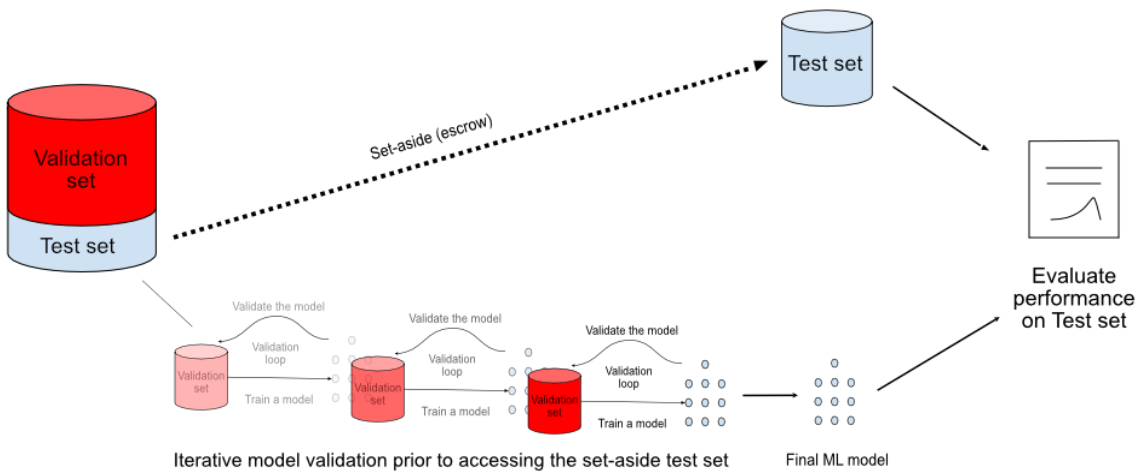

# Tables

(this page intentionally blank - see landscape view in following pages for tables)

Table 1: Five application domains and their perspectives on validation.

|  | ML components / models | Software systems / computerized systems | Medical Device and IVD | Software used in pharmaceutical manufacturing automation | Software used in pharmaceutical drug development (R&D) |
|---|---|---|---|---|---|
| Definition of *validation*. | Use of a sample of data to provide an *unbiased* evaluation of a model fit, on the training dataset, when tuning model hyperparameters. | Confirmation, through the provision of objective evidence, that the *requirements for a specific intended use* or application have been fulfilled (**ISO 9000:2015**)[1]. | Confirmation, through the provision of objective evidence, that the specified users in the specified context of use achieve the specified goals (ie. *intended purpose*). | The **software systems** definition (second column) applies equally here. | The **software systems** definition applies particularly in the *regulatory phase* of development. |

|  | ML components / models | Software systems / computerized systems | Medical Device and IVD | Software used in pharmaceutical manufacturing automation | Software used in pharmaceutical drug development (R&D) |
|---|---|---|---|---|---|
| *What is being validated?*<br><br>i.e. The object or scope of the test | A ML model, trained on a specific set of data (training set) utilizing a specific set of hyperparameters. | Two competing scopes:<br>1. Entire software system (as black box) (narrow validation)<br>2. Entire software system as well as its units and their interactions (broad validation)<br><br>See section, "**Narrow vs broad validation**." | Entire medical device / IVD.<br><br>See section, "**The MD/IVD view on validation.**" | Computerized system in the target hard- and software environment.<br><br>See section, "**Manufacturing automation.**" | Entire software system used to drive a clinical development process including referenced databases.<br><br>See section, "**Drug development (R&D).**" |
| *What is the goal of the validation?*<br><br>i.e. the target of the test | Achievement of most appropriate model hyperparameter selection, and thereby generalizable performance, of model. | Confirmation that software fulfills intended purpose and meets all quality criteria.<br><br>e.g. according to ISO 25010 [25]:<br>- functionality | Confirmation of safety, performance and clinical benefit of the product (narrow validation).<br><br>For IVDs this validation includes clinical performance, | Similar to **software systems** validation (second column).<br><br>Risk based confirmation of conformity with GMP requirements and that software fulfills | Similar to **software systems** validation (second column).<br><br>Risk based confirmation of conformity with GMP requirements and that software fulfills |

|  | ML components / models | Software systems / computerized systems | Medical Device and IVD | Software used in pharmaceutical manufacturing automation | Software used in pharmaceutical drug development (R&D) |
|---|---|---|---|---|---|
|  |  | - performance<br>- compatibility<br>- usability<br>- reliability<br>- security<br>- maintainability<br>- portability | scientific validity, and technical performance. | intended purpose. | intended purpose. |
| *Validation methods.*<br><br>i.e methods for testing validation | Exposing the model to a set-aside, validation-set, containing data which has not been used for training the model.<br><br>Evaluating the performance of the model as expressed by metrics such as sensitivity, specificity, accuracy etc. | From the product life cycle diagram we see examples such as:<br>- software unit tests<br>- integration tests<br>- system tests<br><br>Depending on the test objective these tests include stress and load tests, penetration tests, blackbox-tests etc. | Clinical investigations and evaluations.<br><br>Usability testing. | PQ, IQ, OQ.<br><br>The precise split of responsibilities is defined in GAMP5 - GxP.<br><br>See **Software systems** column for full list of software validation methods applied. | Lab assays, as a form of narrow validation of software performance.<br><br>Software aspects may be reprogrammed from scratch for output comparison (cleanroom implementation).<br><br>See **Software systems** column for full list of software |

|  | ML components / models | Software systems / computerized systems | Medical Device and IVD | Software used in pharmaceutical manufacturing automation | Software used in pharmaceutical drug development (R&D) |
|---|---|---|---|---|---|
|  |  |  |  |  | validation methods applied. |
| Selected legal, best-practices and guidance documents with respect to validation. | Upcoming EU AI regulation [27].<br><br>ISO 24028 [28].<br>ISO/IEC CD TR 29119-11 [29].<br><br>Human-health context:<br>- FDA GMLP guiding principles [30].<br>- AI guideline by Johner Institute [31]<br>- OnRAMP technical best-practices [10]. | See the context specific columns.<br><br>Software used in quality systems:<br>- FDA 21 CFR part 11 (electronic records) [24]<br>- FDA 21 CFR part 820.70 (QS production and process controls) [32]<br>- FDA guidance "software validation" [4].<br>- ISO 13485, e.g. chapter 4.1.6, chapter 7.5.6 [17].<br>- AAMI TIR 36 [33].<br>- ISO/TR 80002-2 | FDA 21 CFR part 820.30 (MD design controls and product validation) [32].<br><br>EU MDR/IVDR Annex I and II [16, 35].<br><br>ISO 13485, chapter 7.3.7 [17]<br><br>IEC 62366-1 [15].<br><br>MEDDEV 2.7/1 rev. 4 [36]. | FDA 21 CFR part 11 (electronic records) [24].<br><br>FDA 21 CFR part 210 and part 211 (GMP) [37, 38]<br><br>EU Directive 2001/83/EC [39].<br><br>GAMP5 - GxP Computerized Systems [18].<br><br>FDA guidance "software validation" [4]. | FDA 21 CFR part 11 (electronic records) [24].<br><br>EU Directive 2001/83/EC [39].<br><br>GAMP5 - GxP Computerized Systems [18].<br><br>FDA guidance "software validation" [4]. |

|  | ML components / models | Software systems / computerized systems | Medical Device and IVD | Software used in pharmaceutical manufacturing automation | Software used in pharmaceutical drug development (R&D) |
|---|---|---|---|---|---|
|  |  | [34]. |  |  |  |
| Typical development phase. | SW component testing, or unit testing (ie. the software unit containing the model). | See section, "***Narrow vs broad validation***." | For narrow validation the focus is at the end of development, before market release. | See section, "***Narrow vs broad validation.***" | Throughout, with particular focus on the 'regulatory phase' where processes may lead to impacts on human health. |